\documentclass[prd,twocolumn,amssymb,amsmath]{revtex4}

\usepackage{graphicx}
\usepackage{txfonts}
\usepackage{aas_macros}

\usepackage{color}

\newcommand{\beq}{\begin{equation}}   %

\newcommand{\eeq}{\end{equation}}   %

\newcommand{\beqa}{\begin{eqnarray}}   %

\newcommand{\eeqa}{\end{eqnarray}}   %

\newcommand{\beal}{\begin{align}}

\newcommand{\enal}{\end{align}}

\newcommand{\bspl}{\begin{split}}

\newcommand{\espl}{\end{split}}

\newcommand{\bsub}{\begin{subequations}}

\newcommand{\esub}{\end{subequations}}

\newcommand{\bmulti}{\begin{multline}}   %

\newcommand{\beqm}{\begin{mathletters}}   %

\newcommand{\eeqm}{\end{mathletters}}   %

\begin{document}
\newcommand{\bx}{{\bf x}}
\newcommand{\bn}{{\bf n}}
\newcommand{\bk}{{\bf k}}
\newcommand{\dd}{{\rm d}}
\newcommand{\dslash}{D\!\!\!\!/}
\def\ga{\mathrel{\raise.3ex\hbox{$>$\kern-.75em\lower1ex\hbox{$\sim$}}}}
\def\la{\mathrel{\raise.3ex\hbox{$<$\kern-.75em\lower1ex\hbox{$\sim$}}}}
\def\beq{\begin{equation}}
\def\eeq{\end{equation}}

\vskip-2cm
\title{\textcolor{black}{Constraints on Gravitino Decay and the Scale of Inflation using CMB spectral distortions}}

\author{{Emanuela Dimastrogiovanni$^a$, Lawrence M. Krauss$^a$$^,$$^b$ and Jens
     Chluba$^c$}}
\affiliation{
     $^a$Department of Physics and School of Earth and Space Exploration, Arizona State University, Tempe, AZ 85827, USA\\
     $^b$Research School of Astronomy and Astrophysics, Mt. Stromlo Observatory, Australian National University,\\ Canberra, Australia 2611\\
      $^c$Institute for Astronomy, K30, University of Cambridge, Madingley Road, Cambridge CB3 0HA, United Kingdom
  }

\vspace*{1cm}

\begin{abstract} 
If local supersymmetry is the correct extension of the standard model of particle physics, then following Inflation the early universe would have been populated by gravitinos produced from scatterings in the hot plasma during reheating. Their abundance is directly related to the magnitude of the reheating temperature. The gravitino lifetime is fixed as a function of its mass, and for gravitinos with lifetimes longer than the age of the universe at redshift $z\simeq 2\times 10^{6}$ (or roughly $6\times 10^6{\rm s}$), decay products can produce spectral distortion of the cosmic microwave background. Currently available COBE/FIRAS limits on spectral distortion can, in certain cases, already be competitive with respect to cosmological constraints from primordial nucleosynthesis for some gravitino decay scenarios. We show how the sensitivity limits on $\mu$ and \textsl{y} distortions that can be reached with current technology would improve constraints and possibly rule out a significant portion of the parameter space for gravitino masses and Inflation reheating temperatures. 

\end{abstract}

\date{\today}

\maketitle

\section{Introduction} 
\label{introduction}

The cosmic microwave background (CMB) temperature and polarization anisotropies represent an invaluable source of information about the origin and evolution of the Universe. They are and have been, for the past few decades, one of the main targets of investigation for cosmology \cite{cmb}.  The CMB, however, presents us with an additional and independent cosmological probe: its energy/frequency spectrum. The frequency spectrum is compatible with a blackbody distribution with an average temperature of 2.726$\,$K \cite{Fixsen:2009ug}. Deviations from a blackbody distribution are potentially generated by any physical process that entails an exchange of energy between matter and radiation \cite{bunch} or the modification of the CMB photon number \cite{bunch_photon}. As such, spectral distortions allow to constrain mechanisms that are within the standard framework of cosmology (including, for instance, recombination \cite{recombination}, reionization and structure formation \cite{reionization}, Silk damping of small-scale fluctuations \cite{damping})  as well as more exotic possibilities, including ones inherent to beyond-the-standard model particle physics (see \cite{exotic} for some examples). \\
\indent At redshifts $z> 2\times 10^{6}$, any produced distortion is quickly erased: double Compton emission, Bremsstrahlung and Compton scattering are efficient enough to immediately restore thermal equilibrium in the primordial plasma. At lower redshifts, $2\times 10^{6}\gtrsim z\gtrsim 5\times 10^{4}$, Compton scattering between photons and electrons is still very rapid whereas double Compton and Bremsstrahlung are no longer efficient. As a result,  a distortion is predominantly produced in the form of a non-vanishing chemical potential ($\mu$ distortion) at high frequencies. Moving down to $z\lesssim 5\times 10^{4}$, Compton scattering also becomes inefficient at restoring kinetic equilibrium and a \textsl{y-type} distortion is created. The latter can be pictured as a high-$z$ version of the Sunyaev-Zeldovish (SZ) effect in galaxy clusters \cite{sz}. During the transition between $\mu$ and y eras, additional (\textsl{r-type}) spectral distortions are produced that cannot be described as a superposition of $\mu$ and \textsl{y} distortions \cite{superposition}. The r-type distortion is crucial if one wants to constrain the time-dependence of phenomena generating spectral distortion around $z\approx 10^{4}-10^{5}$ \cite{rtype}. \\
\indent Current observational bounds on spectral distortions date back to the COBE/FIRAS measurements: these placed upper bounds $|\mu|\lesssim 9\times10^{-5}$ and $|y|\lesssim 1.5\times10^{-5}$ \cite{current}. Modern technology could yield an improvement of more than three orders of magnitude in sensitivity \cite{sensitivity}, a threshold that would allow one to place meaningful bounds on a vast ensemble of processes of relevance for astrophysics and cosmology \cite{added_J}.  \\
\indent Spectral distortions from particle annihilations or decays at $z\lesssim 2\times 10^{6}$ can be used to place constraints on their masses, abundance and interactions \cite{interactions}.  One such particle, the gravitino, is of particular interest and for this reason we focus on gravitinos in this paper.  \\
\indent Gravitinos are spin 3/2 superpartners of the graviton, predicted in the context of supergravity theories (see e.g. \cite{reviewsg} for a review). They are expected to acquire a mass ($m_{3/2}$) via the super-Higgs mechanism. Because of their gravitational strength interactions one might expect that, if inflation \cite{inflation} occurred, any initial abundance of gravitinos would be diluted by the expansion.  Nevertheless, gravitinos can be produced after the end of inflation: thermal production from interactions in the hot plasma during reheating as well as non-thermal effects related to the rapid oscillations of the inflaton, can efficiently replenish the gravitino population \cite{replenish}. Because their interactions are fixed to be of gravitational strength, their post-Inflation abundance is fixed by the reheating scale alone \cite{thermally}. \\
\indent Severe constraints on gravitino abundances arise from cosmology \cite{reviewco}. Gravitinos (or their decay products) surviving until the present time, for instance, would overclose the Universe if their mass density were to exceed the critical density. For unstable gravitinos, decays occurring after big-bang nucleosynthesis (BBN) may ruin the successful predictions of BBN. Decays of gravitinos into neutrinos may affect the effective number ($N_{\text{eff}}$) of relativistic degrees of freedom (see e.g. \cite{Kanzaki:2007pd}), which is also constrained by large scale structure and CMB polarization observations \cite{Abazajian:2013oma}. Because gravitinos lifetime is determined in terms of their mass, sufficiently light gravitinos will decay after $z \approx 10^6$.   If gravitino decays occur during the $\mu$ or the \textsl{y} distortion eras and result into transfer of energy into the photon plasma, they will produce a distortion of the CMB spectrum, in addition to affecting BBN. Both BBN and spectral distortion bounds would result in exclusion regions in the ($T_{\text{rh}}$,$\,m_{3/2}$) plane.   Constraining gravitino abundances in this way can therefore put important new constraints on both the scale of supersymmetry breaking and the scale of Inflation in supersymmetric scenarios.\\
\indent In this paper, we employ current CMB spectral distortion bounds and sensitivity limits for future measurements such as the ones proposed with a PIXIE-like experiment to place upper limits on the reheating temperature, in connection with supersymmetry and the thermal production of unstable gravitinos. This is especially important given that the reheating temperature is otherwise poorly constrained. Reheating only leaves indirect imprints on cosmological observables, which are often dependent on inflationary and reheating model-dependent uncertainties (as e.g. in \cite{reheating}). The only model-independent bounds on $T_{\text{rh}}$ arise from the requirement that reheating should precede BBN \cite{BBN} ($T_{\text{rh}}>1$ MeV). Current bounds on the energy scale of inflation indicate that $T_{\text{rh}}\lesssim \mathcal{O}(10^{16})$ GeV \cite{bicep}, leaving a very large unconstrained parameter space in general. \\
\indent This paper is organized as follows: in Sec.~\ref{overview} we briefly review gravitinos thermal production during reheating along with the implications of an unstable gravitino for BBN; in Sec.~\ref{spectral} we compute the effects of gravitino decay on the CMB frequency spectrum; in Sec.~\ref{conclusions} we offer our conclusions and propose possible future improvements.

\section{Relic gravitinos from reheating} 
\label{overview}

In local supersymmetric theories (supergravity), when SUSY is spontaneously broken, the gravitino acquires a mass by absorbing the Goldstino (Goldstone fermion associated with the broken symmetry). The gravitino mass and the scale ($F$) of SUSY breaking are related by $F\approx \sqrt{m_{3/2}\,M_{\text{P}}}$, where $M_{\text{P}}\approx 2.4\times 10^{18}\,\text{GeV}$ is the Planck mass. \\
\indent During reheating, interactions in the hot plasma lead to gravitino production. The relic density is given by \cite{relic}
\begin{equation}\label{relicc}
n_{3/2}=Y_{3/2} \, s(T)\,,\quad\quad Y_{3/2}\approx 10^{-12}\frac{T_{\text{rh}}}{10^{10}\,\text{GeV}} \,.
\end{equation}
Here $s(T)\equiv (2\pi^{2}/45)g_{*}(T) T^{3}$ is the entropy density of the plasma, with $g_{*}$ the total number of relativistic degrees of freedom. The numerical coefficient in $Y_{3/2}$ can vary depending on the specific value of the cross sections for production processes, but typically results in variations at most of a few in the overall value of the gravitino number density after inflation.\\
The gravitino decay rate is fixed as a function of the mass and of the effective number of decay channels ($N_{\text{dec}}$) \cite{channels}:
\begin{equation}\label{dr}
\Gamma_{3/2}= \frac{N_{\text{dec}}}{(2\pi)}\frac{m^{3}_{3/2}}{M_{\text{P}}^{2}}\,.
\end{equation}    
The beginning of the $\mu$ distortion era ($z\simeq 2\times 10^{6}$) is subsequent to the time frame of BBN (ranging from an initial temperature of $1\,$MeV down to $10\,$keV). The transfer of energy from decaying gravitinos into the CMB photons is most efficient if the decay products are energetic photons \footnote{For general photon injection scenarios, the distortion exhibits a richer phenomenology than the standard $\mu$ and $y$ distortions which are caused by pure heating (see second reference in \cite{bunch_photon}).} or charged particles. As a result, for unstable gravitinos whose decays are relevant to spectral distortion, one also expects important effects on BBN \cite{important}. Predictions of BBN theory for the current abundances of the light elements (mainly D, T, ${}^{3}\text{He}$, ${}^{4}\text{He}$) involve a main parameter, the baryon-to-photon ratio ($\eta_{\text{B}}$).  Agreement between theory and the observed abundances calls for $\eta_{\text{B}}\approx 3\times 10^{-10}$. The outcome of BBN may be entirely different in the presence of gravitinos (or other relic particles decaying after $T\approx 1\,\text{MeV}$). Specifically, there are three main consequences on BBN: (i) the presence of massive gravitinos may affect the expansion rate, leading to an overproduction of ${}^{4}\text{He}$; (ii) radiative decays of gravitinos may lead to a suppression of $\eta_{\text{B}}$; (iii) energetic decay products (such as photons or charged particles) may destroy the light elements. The third class of processes has been shown to represent the dominant effect on BBN for relic particles in the range of masses that we will be concerned with in this work.\\ 
\indent In the next section we will compute the CMB distortion from gravitino decay and present our results along with some constraints from the literature on (iii).

\section{Implications for spectral distortion} 
\label{spectral}

The first step is to compute the rate of energy release from gravitino decay
\begin{equation}\label{sp10}
\frac{\text{d}E_{\text{dec}}}{\text{dt}}=\epsilon_{3/2} m_{3/2} \frac{1}{a(t)^{3}}\frac{d}{dt}\left[a(t)^{3}N_{3/2}(t)\right]\,.
\end{equation}
Here $\epsilon_{3/2}$ is a dimensionless parameter quantifying the fraction of energy from the decay products that contributes to heating of the CMB photon bath via Comptonization, $m_{3/2}$ is the gravitino mass, $a$ the scale factor and $N_{3/2}$ is the number density of gravitinos. Following the parametrization in \cite{parametrization}, Eq.~(\ref{sp10}) can be rewritten as 
\begin{equation}\label{eq1}
\left\|\frac{\text{d}E_{\text{dec}}}{\text{dt}}\right\|  =f_{3/2}\,N_{\text{H}}\,\Gamma_{3/2}\,e^{-\Gamma_{3/2} t}\,,
\end{equation}
where the quantity $f_{3/2}$ collects all of the information about the decaying particle (its mass/lifetime as well as its abundance) and about the decay process (e.g. the number of channels or the number of effective relativistic degrees of freedom produced from the decay). In Eq.~(\ref{eq1}), $N_{\text{H}}$ is the number density of hydrogen nuclei and $\Gamma_{3/2}$ the decay rate of gravitinos. It can be shown that $f_{3/2}$ (or $f_{3/2}/z_{3/2}$, where $z_{3/2}$ indicates the redshift at which the decay occurs) factors out of the integrals in the definitions of the effective $\mu$ and \textsl{y} distortion parameters, thus proving a very convenient choice for parametrizing spectral distortion signals from decaying particles. With the above definitions, and using Eqs.~(\ref{relicc}) and (\ref{dr}), one arrives at 
\begin{equation} \label{newenergy}
\frac{f_{3/2}}{z_{3/2}}\simeq \epsilon_{3/2}\, \frac{10^{-6}}{N_{\text{dec}}^{1/2}}\,\left(\frac{T_{\text{rh}}}{\text{GeV}}\right)\left(\frac{m_{3/2}}{\text{GeV}}\right)^{-1/2} \text{eV} \,,
\end{equation}
where we used $N_{\text{H}}\approx 1.9\times 10^{-7}\,(1+z)^{3}\,\text{cm}^{-3}$ for the number density of hydrogen atoms. The relation between gravitino mass and lifetime is $t_{3/2}\simeq (2.4\times 10^{13}/N_{\text{dec}})\,\left(m_{3/2}/\text{GeV}\right)^{-3} \,\text{s}$. We use a conventional  time-temperature relation, $t\approx[\sqrt{45}\,M_{\text{P}}]/[\sqrt{2\pi^{2}\,g_{*}(T)}T^{2}]$, and time-redshift relation, $z\approx 4.9\times 10^{9}/\sqrt{t/{\rm s}}$ throughout.\\
\indent The $r$-distortion appearing at intermediate redshifts ($10^{4}\lesssim z\lesssim \text{few} \times 10^{5} $) normally requires a numerical treatment \cite{superposition}. In this paper, we wish to retain analytic control over our calculations. We will therefore focus on $\mu$ and \textsl{y} distortions, for which simple analytic approximations have been found, leaving the study of any intermediate-type distortion for future work. It is also well known that, at smaller redshifts, another contribution to the y-distortion arises, namely that due to the inverse Compton scattering of CMB photons off free electrons (thermal Sunyaev-Zeldovich, tSZ, effect) \cite{reionization, tSZ}. Thus, limits derived from the $y$-parameter should be interpreted as conservative upper limits.\\
\indent We write the fractional variation of the energy density of CMB photons as the sum of $\mu$ and \textsl{y} distortion contributions 
\begin{equation}\label{ref_2}
\frac{\Delta \rho_{\gamma}}{\rho_{\gamma}}\approx\left[\frac{\Delta \rho_{\gamma}}{\rho_{\gamma}}\right]_{\mu}+\left[\frac{\Delta \rho_{\gamma}}{\rho_{\gamma}}\right]_{y}\,,
\end{equation}
where the effective distortion parameters are given by \cite{effectived}
\begin{eqnarray}
\left[\frac{\Delta \rho_{\gamma}}{\rho_{\gamma}}\right]_{\mu}\equiv\frac{\mu}{1.401}\,,\quad\quad\left[\frac{\Delta \rho_{\gamma}}{\rho_{\gamma}}\right]_{y}\equiv 4\,y\,.
\end{eqnarray}
For pure $\mu$ and \textsl{y} distortion one has  
\begin{equation}\label{ref__1}
\mu\approx 1.4 \int\mathcal{J}_{\rm bb}\,\mathcal{J}_{\mu} \frac{1}{\rho_{\gamma}}  \left(\frac{\text{d}E}{\text{dt}}\right)\,dt\,,\,\,\, 
y\approx \frac{1}{4} \int \mathcal{J}_{\rm bb}\,\mathcal{J}_{y} \frac{1}{\rho_{\gamma}}  \left(\frac{\text{d}E}{\text{dt}}\right)\,dt \,.
\end{equation}
where the thermal response of the medium to the energy injection has been parametrized with the \textsl{visibility functions} \cite{visibility}
\begin{eqnarray}
&&\mathcal{J}_{\rm bb}(z)\approx  \exp\left[-(z/z_{\mu})^{5/2}\right] \,,\\
&&\mathcal{J}_{y}(z)\approx \left[1+\left(\frac{1+z}{6.0\times 10^{4}}\right)^{2.58}\right]^{-1},\,\,\, \mathcal{J}_{\mu}(z)\approx 1-\mathcal{J}_{y}(z)\,.\nonumber
\end{eqnarray}
\begin{figure}
\centering
\includegraphics[width=1.0\columnwidth]{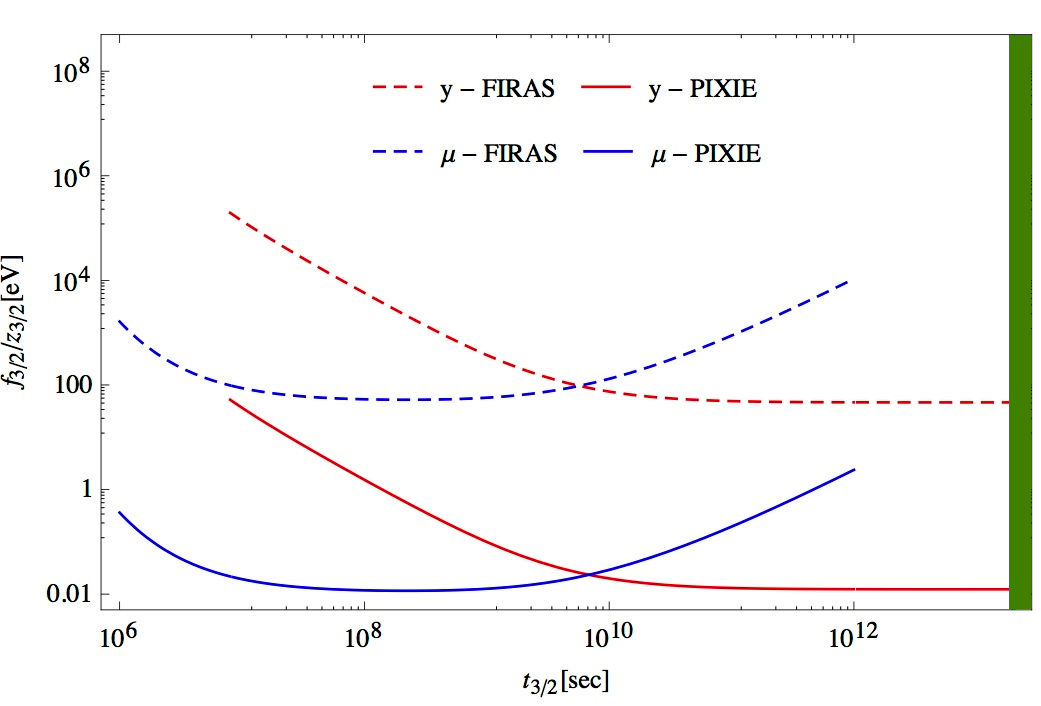}
\caption{Limits placed by FIRAS (dashed lines) and sensitivity projections for PIXIE (solid lines) from $y$ and $\mu$ distortions on the effective energy input per hydrogen atom normalized to the redshift at decay, $f_{3/2}/ z_{3/2}$, and the lifetime, $t_{3/2}$. The range of values of $t_{3/2}$ encompasses the whole $\mu$-distortion era and the \textsl{y} distortion-era until around recombination (green band). The areas above the dashed lines have been excluded by FIRAS. }
\label{fig1}
\end{figure}
Here $\mathcal{J}_{\rm bb}$ accounts for the fact that thermal equilibration processes are highly efficient at $z_{\mu}\approx 2\times 10^{6}$. The definition of $ \mathcal{J}_{\mu}$ follows from enforcing energy conservation according to Eq.~(\ref{ref_2}) and $\mathcal{J}_{y}$ was found to approximate the branching of energy eventually appearing as $y$-distortion \citep{visibility}. In Eq.~(\ref{ref__1}), we set the lower bound in redshift in the integral for y-distortion to $z\approx 1000$. The y-distortion is in principle produced down to $z\approx 200$ (at $z<200$ the rate of baryons-photons interactions becomes too low for thermodynamic equilibrium to be maintained). However, at $z<10^3$ a more detailed treatment of the energy exchange between matter and radiation may be required as the plasma recombines. \\
\indent From the observed limits on $\mu$ and $y$ from COBE/FIRAS and from the forecast sensitivity of an experiment like PIXIE ($|\mu|\lesssim2\times10^{-8}$ and $|y|\lesssim 4\times10^{-9}$), one obtains the bounds in Fig.~\ref{fig1} in the $(f_{3/2}/z_{3/2}\,,t_{3/2})$ plane. Red lines are from $y$, blue lines from $\mu$ distortions. Dashed lines denote the lower contours of the exclusion regions defined by FIRAS observations. Solid lines show the sensitivity limits of PIXIE.  \\
\begin{figure}
\centering
\includegraphics[width=0.99\columnwidth]{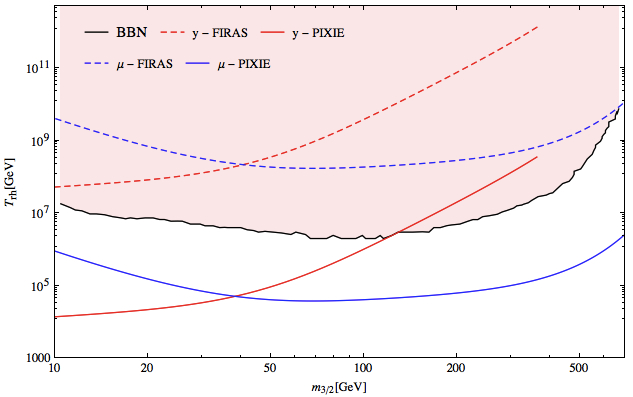}
\\[2mm]
\includegraphics[width=0.99\columnwidth]{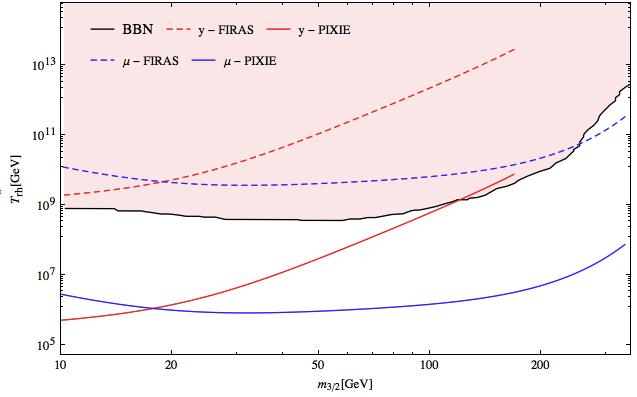}
\\[2mm]
\includegraphics[width=0.99\columnwidth]{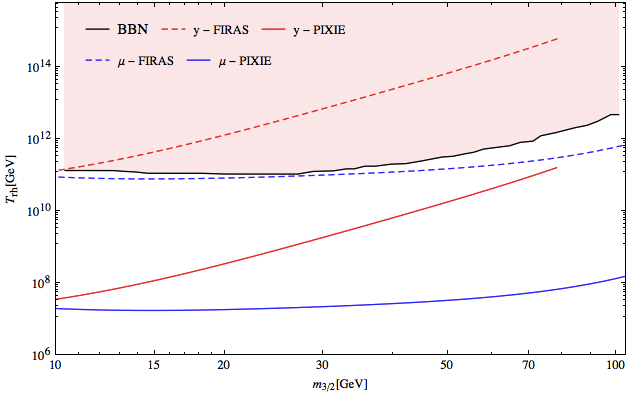}
\caption{Collection of FIRAS and BBN exclusion regions in the reheating temperature - gravitino mass plane, along with PIXIE sensitivity limits (solid lines). The shaded area is ruled out by BBN \cite{carried1}, dashed lines define the lower boundaries of the areas ruled out by FIRAS. Color codes are as in Fig.~\ref{fig1} for y and $\mu$ distortions. The energy release scenario is that of a decay of a gravitino into a photon $+$ photino. The three figures correspond to decreasing values of the branching ratio for this process: $B_{[\text{G}\rightarrow \gamma+\tilde{\gamma}]}=1,\,0.1,\,0.01$, respectively moving from the figure at the top to the one at the bottom. }
\label{fig2}
\end{figure}
\indent The joint bounds on $T_{\text{rh}}$ and $m_{3/2}$ are derived from the ones on ($f_{3/2}/z_{3/2},\,t_{3/2}$) 
\begin{eqnarray} \label{temperature}
&&\frac{T_{\text{rh}}}{\text{GeV}}\simeq \left(\frac{f_{3/2}}{z_{3/2}\,\text{eV}}\right)\frac{10^{6}\,N_{\text{dec}}^{1/2}}{\epsilon_{3/2}}\left(\frac{m_{3/2}}{\text{GeV}}\right)^{1/2} \,,\\\label{mass}
&&\frac{m_{3/2}}{\text{GeV}}\simeq  \left(\frac{2.4 \times 10^{13}}{N_{\text{dec}}^{}}\right)^{1/3}\left(\frac{t_{3/2}}{\text{s}}\right)^{-1/3} \,.
\end{eqnarray}
For given values of the gravitino mass, $N_{\text{dec}}$ and $\epsilon_{3/2}$, the temperature given in Eq.~(\ref{temperature}) represents: (i) the maximum reheating temperature currently allowed by FIRAS y and $\mu$ distortion bounds, for $f_{3/2}/z_{3/2}$ given by the FIRAS lines in Fig.~\ref{fig1}; (ii) the smallest value for the upper bound that an experiment like PIXIE would be able to place on the reheating temperature, for $f_{3/2}/z_{3/2}$ given by the PIXIE lines in Fig.~\ref{fig1}. Notice that, as one approaches $\epsilon_{3/2}\rightarrow 0$ (limit of no energy transfer to the photons), $T_{\text{rh}}$ in Eq.~(\ref{temperature}) becomes larger, i.e. the constraint from spectral distortion weakens. The other extreme is $\epsilon_{3/2}\rightarrow 1$, which provides the most stringent bounds that can be obtained for a given process.\\
\indent Gravitinos decay through a variety of channels. Several studies have been proposed that quantify the effects of the decay on the BBN predictions for light elements primordial abundances \cite{important}. We will refer to the studies carried out in \cite{carried1} and \cite{carried2} and include some of their results in our plots, alongside our spectral distortion sensitivity lines, so as to visualize the different cosmological constraints simultaneously. \\
\indent Unsurprisingly, both in the context of BBN and of spectral distortion the most severely constrained scenarios involve direct decays into charged particles and/or into photons. \\
\indent We will first consider the case of a gravitino decaying into photon $+$ photino. If the photino mass is $m_{\tilde{\gamma}}\ll m_{3/2}$, the total decay rate reads 
\begin{equation}\label{assuming}
\Gamma_{3/2}\simeq\frac{m_{3/2}^{3}}{32\,\pi\,M_{\text{P}}^{2}}\,.
\end{equation}
Our results for this case are shown in the first plot of Fig.~\ref{fig2}. Here we set $\epsilon_{3/2}=1/2$: the fraction of gravitino initial energy effectively converted into photons is the one ultimately responsible for CMB distortion. \\
\begin{figure}
\centering
\includegraphics[width=1.0\columnwidth]{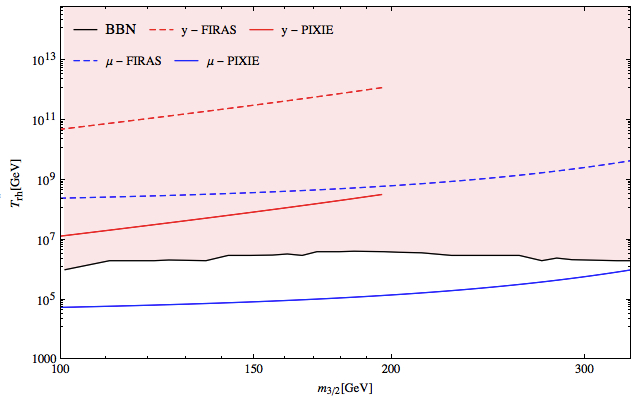}
\caption{Limits from spectral distortion (blue and red lines) and from BBN (black line, reproducing results in \cite{carried2}) for gravitinos decaying entirely into hadrons. }
\label{fig5}
\end{figure}
\indent The effect of such a decay on BBN are well-studied. Photons emitted during the decay initiate an electromagnetic cascade. They can scatter off background photons transferring energy to the latter or producing, for example, electron-positron pairs. Photons can also interact with matter, scattering with background electrons or producing pair creation in the presence of nuclei.\\
\indent We consider the results obtained in \cite{carried1}; here the spectrum of high energy photons and electrons was computed and from it photodissociation effects on the light elements were quantified. The BBN lines shown in Fig.~\ref{fig2} arise primarily from D and ${}^{3}\text{He}$ abundances: the shaded region corresponds to an overproduction of these elements and is therefore excluded. Notice that the bounds from BBN are at least one order of magnitude stronger than the FIRAS limits in most of the mass range reported in the figure, with spectral distortion limits approaching the ones from BBN only near the extreme ends of the range, i.e. around $m_{3/2}\approx 10\,\text{GeV}$ and towards $m_{3/2}\approx 700\,\text{GeV}$. The latter value corresponds to gravitinos decaying at the onset of the $\mu$ era. In this plot (the same will apply to the remaining plots of Fig.~\ref{fig2}), in order to draw a comparison with the limits from BBN studies in the literature, we considered $m_{3/2}\approx 10\,\text{GeV}$ as the lowest value of our mass range. However, our spectral distortion bounds also cover the $\mathcal{O}(1-10)$ GeV range, which we show in Fig.~\ref{fig6} (upper panel). \\
\indent Our plot shows that an experiment like PIXIE has the potential to provide highly competitive constraints on the reheating temperature: for masses $10\,\text{GeV}\lesssim m_{3/2}\lesssim 100\,\text{GeV}$, $T_{\text{rh}}^{\text{max}}\approx 10^{6}\,\text{GeV}$ from BBN, whereas PIXIE may be able to probe reheating temperatures as low as $6\times 10^{3}\,\text{GeV}$.

\begin{figure}
\centering
\includegraphics[width=0.99\columnwidth]{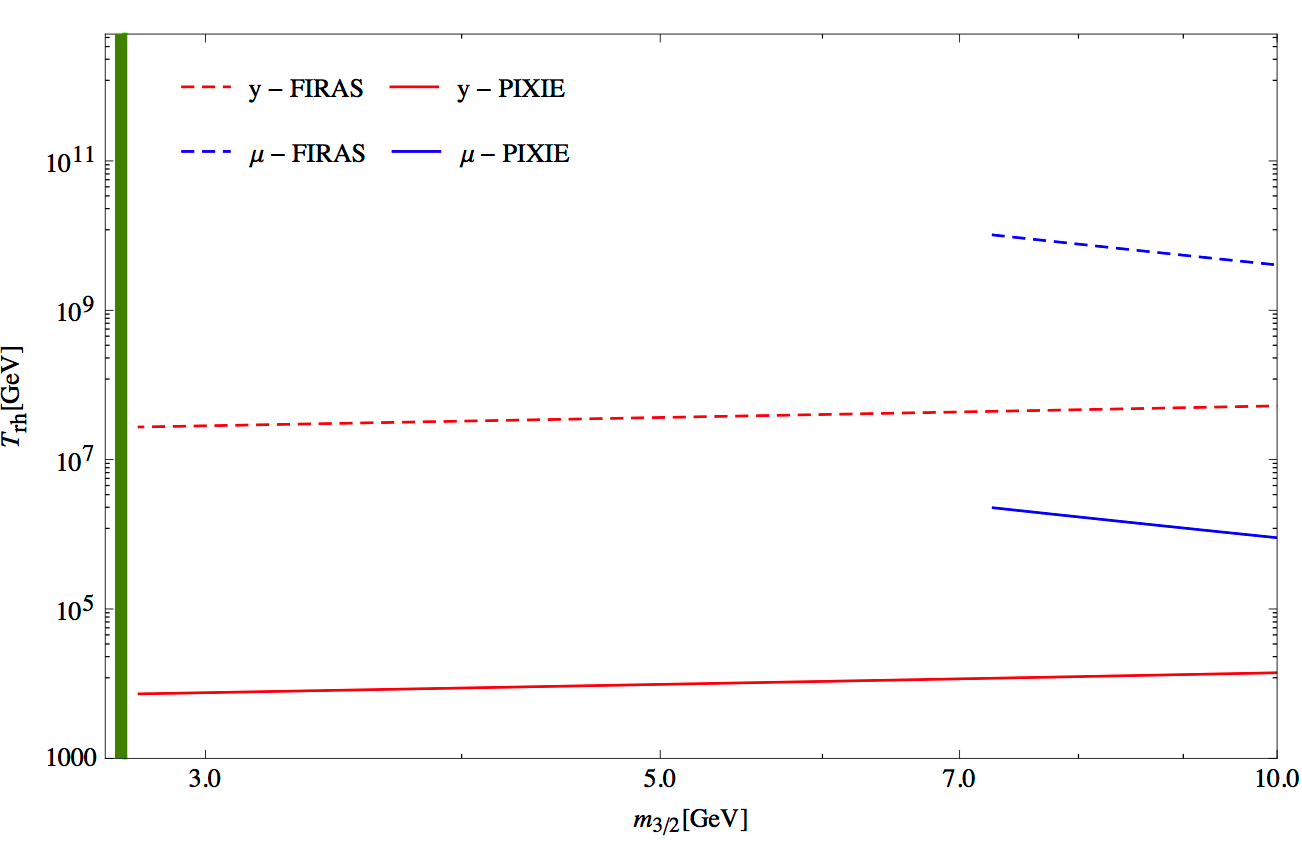}
\\[2mm]
\includegraphics[width=0.99\columnwidth]{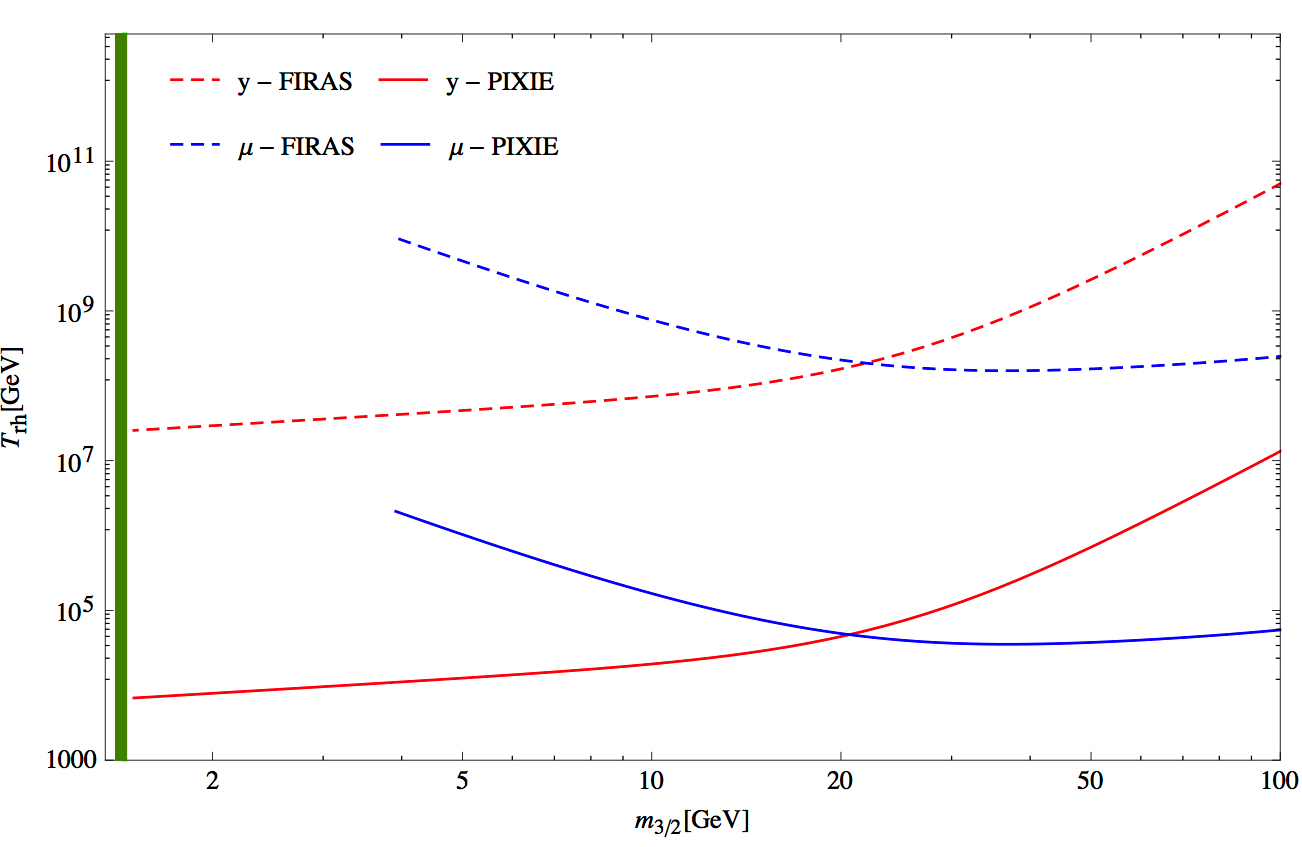}
\caption{FIRAS exclusion regions (dashed lines) and PIXIE sensitivity limits (solid lines) for gravitinos decaying before recombination (marked by the green band). Color codes are as in Fig.~\ref{fig1} for y and $\mu$ distortions. The energy release scenario is that of a decay of a gravitino into photon $+$ photino (upper panel) and of a decay into hadrons (lower panel), both with unitary branching ratios. }
\label{fig6}
\end{figure}

\indent In supergravity models where R-parity is preserved, the lightest supersymmetric particle (in our case the photino) is stable. One then should also require that the energy density of relic photinos does not overcome the critical density today. This results in an additional bound on the reheating temperature \cite{carried1}: $T_{\text{rh}}\lesssim 10^{11}\,(m_{\tilde{\gamma}}/100\,\text{GeV})^{-1} h^{2}\,\text{GeV}$, where $h$ is the Hubble rate in units of $100\,(\text{km}/\text{sec})/\text{Mpc}$. The photino mass is strictly model-dependent. If $m_{3/2}$ is viewed as an upper bounds for $m_{\tilde{\gamma}}$ (condition for Eq.~(\ref{assuming}) to apply), then in the above range for the gravitino mass, the bound derived on the reheating temperature from the relic density of photinos is much weaker than both BBN and spectral distortion bounds.\\
\indent If gravitinos only partially decay into photons and photinos, as will in general be the case, one may describe this by introducing an additional parameter, the branching ratio $B_{[\text{G}\rightarrow \gamma+\tilde{\gamma}]}\equiv \Gamma_{[\text{G}\rightarrow \gamma+\tilde{\gamma}]}/\Gamma_{\text{total}}$. A value $B_{[\text{G}\rightarrow \gamma+\tilde{\gamma}]}=1$ would then correspond to the results just discussed and represented in the upper panel of Fig.~\ref{fig2}. BBN bounds can be derived for different values of $B_{[\text{G}\rightarrow \gamma+\tilde{\gamma}]}$ in \cite{carried1}. Being blind to the effect of the remaining decay channels, some of which will likely have cosmological implications, the bounds derived with this procedure will therefore be conservative.   For the sake of comparison between spectral distortion and BBN constraints, we adopt this simplified approach here. For consistency we assume that the ratio of the initial gravitino energy that is transferred to the CMB bath is simply reduced by a factor equal to the branching ratio w.r.t. the case where gravitinos entirely decay into photons and photino, i.e. $\epsilon_{3/2}=B_{[\text{G}\rightarrow \gamma+\tilde{\gamma}]}/2$. \\
\indent Our results are represented in the second and third panels of Fig.~\ref{fig2}, respectively for $B_{[\text{G}\rightarrow \gamma+\tilde{\gamma}]}=0.1$ and $B_{[\text{G}\rightarrow \gamma+\tilde{\gamma}]}=0.01$.  Notice that for these more realistic scenarios, the $\mu$ distortion bounds from FIRAS are now comparable to or slightly stronger than BBN bounds for the largest plotted values of the gravitino mass in the $B_{[\text{G}\rightarrow \gamma+\tilde{\gamma}]}=0.1$ case, and in the whole mass range for $B_{[\text{G}\rightarrow \gamma+\tilde{\gamma}]}=0.01$. As for $B_{[\text{G}\rightarrow \gamma+\tilde{\gamma}]}=1$, our plots show that also for these smaller values of the branching ratio PIXIE would be able to rule out a substantial portion of the currently allowed parameter space. One could gain access to temperature values of $T_{\text{rh}}$ down to $10^{5}-10^{6}$ GeV, as opposed to the $T_{\text{rh}}\gtrsim 10^{9}-10^{11}$ GeV range one can probe with BBN and current spectral distortion bounds for gravitino masses $10\,\text{GeV}\lesssim m_{3/2}\lesssim 100-300\,\text{GeV}$.  \\
\indent We can also consider exclusively hadronic decay channels for gravitinos, where one finds 
\begin{equation}
\Gamma_{3/2}\approx \frac{m_{3/2}^{3}}{5\,\pi\,M_{\text{P}}^{2}}\,.
\end{equation}
We refer to the nucleosynthesis bounds obtained for this case in \cite{carried2}, which were derived for a gravitino with mass $m_{3/2}\gtrsim 100\,\text{GeV}$. The heaviest gravitinos that $\mu$ distortion can constrain in this case have masses $m_{3/2}\lesssim 300\,\text{GeV}$. In Fig.~\ref{fig5} we present our spectral distortion results (for which we set $\epsilon_{3/2}\approx 1$) and also display the primordial nucleosynthesis bounds from \cite{carried2} (mostly due to ${}^{3}\text{He}/\text{D}$ and to ${}^{6}\text{Li}/\text{H}$ measurements) in the overlapping mass range. The $\mu$ distortion bounds from an experiment like PIXIE would be more stringent than nucleosynthesis bounds in most of the mass range, nearing BBN for the heaviest mass values. In Fig.~\ref{fig6} (lower panel) we present the FIRAS exclusion regions and the PIXIE sensitivity bounds for the range $\mathcal{O}(1-100)\,\text{GeV}$ of gravitino masses.\\
\indent  Our results in Eqs.~(\ref{newenergy}), (\ref{temperature}) and (\ref{mass}) are completely general and therefore applicable to any scenario for gravitino decay, simply by varying $N_{\text{dec}}$ and $\epsilon_{3/2}$. We should also mention that the spectral distortion constraints are derived assuming that only a small fraction of all the energy transferred to the medium by the decay products is absorbed by light elements, and also ignoring a variety of other possible energy injection mechanisms beyond decay into photons. A more complete analysis should simultaneously follow the effective fraction of energy used up by destroying light elements, and also more complete energy injection cascades.  The former effect might at best slightly weaken our bounds.  The latter is likely to strengthen them. We leave such an analysis to a future paper.

\section{Conclusions and outlook} 
\label{conclusions}

Cosmological datasets offer various routes to uncovering beyond-the-standard model particle physics. Supersymmetry is a promising candidate for extending the standard model and searching for SUSY-induced effects is one of the main goals in collider experiments. If SUSY is the correct description of nature, it must broken at some energy scale. We do not know what the SUSY breaking mechanism is, nor the scale at which it occurs. Local supersymmetry (supergravity) theories predict the existence of the gravitino, spin $3/2$ superpartner of the graviton. The gravitino mass is related to the SUSY breaking scale and its interactions are fixed in a nearly model-independent way. Being able to obtain stringent constraints on gravitinos is then invaluable for testing supergravity. \\
\indent Moreover, in the early Universe, gravitinos will be generated thermally from interactions in the thermal bath during reheating following Inflation. In this case, their abundance would be a function of the reheating temperature. If gravitinos decay after $z\approx 2\times10^{6}$, they may produce observable distortions of the cosmic microwave background frequency spectrum. Thus constraints on gravitino decays can provide important constraints on the scale of Inflation.\\
\indent In this paper we have analyzed how, with current technology, spectral distortions in the CMB constrain a sizable region of the reheating temperature - gravitino mass parameter space.  We have derived general analytic expressions for computing $\mu$ and \textsl{y} distortions from gravitino decays occurring between the beginning of the $\mu$ distortion era and recombination. Our results are expressed in terms of a few parameters describing the number and types of decay channels. \\
\indent We have plotted the exclusion regions in ($T_{\text{rh}},\,m_{\text{3/2}}$) space for COBE/FIRAS along with the sensitivity limits of a PIXIE-like experiment for various simplified assumptions regarding gravitino decay, considering energy injection purely by direct photon or hadron decay products. We show that, when compared with the bounds from primordial nucleosynthesis, a PIXIE-like experiment would be able to constrain a much larger region of parameter space and that the bounds from FIRAS can be competitive or exceed those derived from BBN considerations (see Fig.~\ref{fig2} and Fig.~\ref{fig5}).  We find that a PIXIE-like experiment will be able to constrain inflationary reheating temperatures as low as $6 \times 10^3$ GeV, which will cover most of the allowed parameter range for inflation, and therefore interestingly might allow a detection of SUSY-related Inflation, rather than a simple constraint on models. \\
\indent Our study can be extended in several ways. It would be interesting to also include $r$-distortion, by performing a numerical analysis of the thermal response of CMB photons to gravitino decay. We also limited our study to gravitinos decaying before recombination. Numerical tools have been developed for extending our work to include lighter gravitino masses corresponding to later decays. These are predicted in a large number of supersymmetric models and are far from being ruled out by other cosmological probes.  Finally, more stringent bounds could likely be derived with calculations of full charged particle cascades beyond simple initial photon or hadron decay products. As our analysis suggests, such improvements are worth considering, given the great reach of CMB spectral distortions for constraining gravitino physics and the physics of Inflation.\\ 

\section*{Acknowledgments}

E.D. is grateful to Yue Zhao for fruitful discussions. J.C. is supported by the Royal Society as a Royal Society University Research Fellow at the University of Cambridge, UK.  E.D. and L.M.K. acknowledge support from the DOE under grant No. DE-SC0008016.




\end{document}